\documentclass[11pt]{article}

\usepackage[utf8]{inputenc}
\usepackage[top=1in,bottom=1in,left=0.75in,right=1in]{geometry}
\usepackage{amssymb,amsmath}
\usepackage{feynmp}
\usepackage[force]{feynmp-auto}
\usepackage{esvect}
\usepackage{slashed}
\usepackage{graphicx}
\usepackage{dsfont}

\newcommand{\be}{\begin{eqnarray}}
\newcommand{\ee}{\end{eqnarray}}
\newcommand{\id}{{\mathbb I}}
\newcommand{\im}{{\rm i}}

\begin{document}
\begin{fmffile}{diagram}

\title{One-loop same helicity four-point amplitude from shifts}
\author{Pratik Chattopadhyay and Kirill Krasnov\\ {}\\
{\it School of Mathematical Sciences, University of Nottingham, NG7 2RD, UK}}

\date{February 2020}
\maketitle
\begin{abstract}It has been suggested a long time ago by W. Bardeen that non-vanishing of the one-loop same helicity YM amplitudes, in particular such an amplitude at four points, should be interpreted as an anomaly. However, the available derivations of these amplitudes are rather far from supporting this interpretation in that they share no similarity whatsoever with the standard triangle diagram chiral anomaly calculation. We provide a new computation of the same helicity four-point amplitude by a method designed to mimic the chiral anomaly derivation. This is done by using the momentum conservation to rewrite the logarithmically divergent four-point amplitude as a sum of linearly and then quadratically divergent integrals. These integrals are then seen to vanish after appropriate shifts of the loop momentum integration variable. The amplitude thus gets related to shifts, and these are computed in the standard textbook way. We thus reproduce the usual result but by a method which greatly strengthens the case for an anomaly interpretation of these amplitudes. 
\end{abstract}

\section{Introduction}

The same helicity Yang-Mills (YM) amplitudes vanish at tree-level, but become non-zero at one-loop. The QCD one-loop amplitudes at four (and five) points were computed by the field theory techniques in \cite{Ellis:1985er}, and via string-inspired technology in \cite{Bern:1991aq} (four-points) and \cite{Bern:1993mq} (five-points). The result for same helicity five gluon amplitude was then used to conjecture \cite{Bern:1993sx} an $n$-gluon formula. Supersymmetry implies that there is a relation between same helicity one-loop amplitudes in theories with different spin particles (e.g. spin 1 and spin $1/2$) propagating in the loop, see \cite{Bern:1993mq}. This means that the same helicity one-loop amplitudes in YM are related to those in massless QED. The later were computed in  \cite{Mahlon:1993fe} using recursive methods, thus proving the conjecture of \cite{Bern:1993sx}. This conjecture received additional support from the consideration of the collinear limits in \cite{Bern:1993qk}. At four points, which is the case of main interest for us in this paper, the same helicity amplitude takes the following extremely simple form \cite{Bern:1991aq}
\be\label{amplitude-intr}
{\cal A}_{\rm one-loop}(1^+,2^+,3^+,4^+)\sim \frac{[12][34]}{\langle 12\rangle \langle 34\rangle},
\ee
where the spinor helicity notations are used, see below, and the proportionality factor contains a numerical coefficient as well as powers of the relevant coupling constant. 

In \cite{Bardeen:1995gk} William A. Bardeen suggested that the integrability of the self-dual sector of YM theory is behind the simplicity of the all-same helicity sector of the full YM. This paper also conjectured that the non-vanishing of the same-helicity one-loop amplitudes should be interpreted as the anomaly of the currents responsible for the integrability of the self-dual sector. The paper \cite{Cangemi:1996rx} explicitly confirmed that the four-point same helicity one-loop amplitudes in self-dual YM are given by (\ref{amplitude-intr}). This paper also discusses symmetries of the self-dual YM.

More recently, the non-vanishing of the one-loop same helicity amplitude in YM and gravity was shown to be linked to the UV divergence of the two-loop quantum gravity. This is very clear from the calculations \cite{Bern:2015xsa}, \cite{Bern:2017puu} that use unitarity methods and directly link the two-loop divergence to the non-vanishing of the same helicity one-loop amplitude. It thus becomes more pressing to revisit the possible anomaly interpretation of the one-loop same helicity amplitude. Indeed, if this amplitude's non-vanishing is the signal of an anomaly, it may be made to vanish by appropriately canceling the anomaly. It is thus very important to understand the anomaly interpretation, if any, of the same helicity one-loop amplitudes of YM and gravity. 

The purpose of this paper is a modest step in this direction. The available calculations \cite{Bern:1991aq}, \cite{Mahlon:1993fe}  of the four-point one-loop same helicity amplitude are not transparent. The first of these uses string theory inspired methods. The second calculates all $n$-point amplitudes (in massless QED) and uses usual Feynman diagrams but resorts to dimensional regularisation to extract the final result. Given that the amplitude one calculates is non-divergent, this makes it hard to understand where the result is coming from. All this makes it very difficult to understand how such a simple answer as (\ref{amplitude-intr}) that consists of a single rational term can arise from a one-loop computation. Thus, the result itself does strongly suggest the anomaly interpretation, but available derivations of this result are nowhere near the simplicity of e.g. the textbook derivation of the chiral anomaly. 

The purpose of this paper is to provide a new computation of the result (\ref{amplitude-intr}) by a method that mimics the textbook triangle diagram chiral anomaly computation. In the latter case, the linearly divergent triangle diagram can be rewritten as a sum of quadratically divergent integrals, all of which can be seen to either vanish directly or vanish after appropriate shifts of the integration variable. More specifically, consider the quadratically divergent one-loop integral of the type
\be\label{integr-intr}
\int \frac{d^4l}{(2\pi)^4}\frac{l_\mu l_\nu}{l^2(l-p)^2},
\ee
where $l$ is the loop momentum and $p$ is some external momentum. Then, to argue that the integral of this sort vanishes (when contracted with other objects), we use Lorentz invariance to conclude that it can only be proportional to $p^2 \eta_{\mu\nu}$ or $p_\mu p_\nu$. This argument is a standard textbook one, see e.g. \cite{Srednicki:2007qs}, formula (75.27) for similar reasoning. Below we will review how the chiral anomaly calculation can be reduced to a computation of shifts by the argument of this type. 

The method that we use to compute the same helicity one-loop amplitude is completely analogous. We use the momentum conservation to rewrite the nominally logarithmically divergent integral as a sum of linearly divergent integrals, and then use the momentum conservation once more to rewrite everything as a linear combination of quadratically divergent integrals. At this step we can see that the integrals are either directly of the type (\ref{integr-intr}) and so vanish when contracted with other objects, or become integrals of this type if the loop momentum integration variable can be shifted. The amplitude thus gets related to shifts, which are computed in the standard way. This makes the computation of the one-loop same helicity amplitude resemble the chiral anomaly computation, and thus strongly supports the anomaly interpretation of the result (\ref{amplitude-intr}). 

We do not aim to answer the "anomaly of what" question in this paper. Nevertheless, Bardeen's suggestion \cite{Bardeen:1995gk} that the amplitude (\ref{amplitude-intr}) as well as its higher point counterparts signal the anomalous non-conservation of the currents responsible for the integrability of the self-dual YM still stands. It is just that more work is needed to demonstrate this convincingly. 

We perform all computations for massless QED, but in Appendix we review the version of the Feynman rules for self-dual YM (SDYM) that makes it manifest that the SDYM amplitudes are multiples of those in massless QED. The case of gravity is not treated in this paper and is left to future work, but we expect that a similar computation is also possible there. 

The organisation of the rest of this paper is as follows. Next Section contains a warm up and rewrites the textbook chiral anomaly computation using the 2-component spinor formalism. Essentially, the purpose here is to fix notations and to also illustrate the index-free spinor notation that we use throughout. The computation of real interest is contained in Section \ref{sec:box}. Some technical results, and in particular the Feynman rules that we use, are contained in the Appendices. 

\section{Triangle anomaly}
\label{sec:triangle}

All computations in this paper are performed for massless QED, as the Feynman rules are simplest in this case. The case of self-dual YM can be treated completely analogously, and we discuss the (minimal) necessary changes in the Appendix, where we present the corresponding Feynman rules. 

We use 2-component spinors throughout, and our 2-component spinor conventions are explained in e.g. the Appendix of \cite{Krasnov:2016emc}. We adopt spinor index-free notation of bras and kets. Briefly, a 4-momentum $l_\mu$, where $\mu$ is the usual spacetime index, becomes the spinorial object $l_M{}^{M'}$. It can thus "act" on a primed spinor $\lambda_{M'}$ returning an object $l_M{}^{M'} \lambda_{M'}$  that is an unprimed spinor. In index-free notation the primed spinor $\lambda_{M'}$ is represented by a ket $|\lambda]$, where the square bracket indicates that this is a primed spinor. The umprimed spinor $l_M{}^{M'} \lambda_{M'}$ gets represented as the result of action of $l$ on the ket $|\lambda]$, which we write as $l|\lambda]$. This can now be contracted with an arbitrary unprimed spinor $\mu^M$. Our convention is that all kets are 2-component spinors with indices in the lower position. The bras are 2-component spinors with their spinor indices in the upper position. So, $\mu^M l_M{}^{M'} \lambda_{M'}$ is, in index-free notations, $\langle \mu| l| \lambda]$. 

Other convenient conventions are as follows. All external momenta are assumed to be incoming. We label external momenta on any diagram as $k_1, k_2, \ldots$. But it is very convenient, when no confusion can arise, to drop the letter $k$ and refer to the momentum just by its number. Thus, our standard notation is $k_{1\, \mu} \equiv 1_\mu$, which in spinor notations becomes $1_M{}^{M'}$. When this momentum is null, then $1_M{}^{M'}= 1_M 1^{M'}$. In index-free notations this becomes
\be\label{k1}
k_1 \equiv 1 = |1\rangle [1|.
\ee
We use the spinor helicity formalism throughout, and a positive helicity state that corresponds to a momentum e.g. 2 is given by $\epsilon^+_{2\, M'}{}^M = 2_{M'} q^M/ \langle q2\rangle$, where $\langle q2\rangle:= q^M 2_M$. In index-free notations such a state becomes 
\be
\epsilon^+_2 = \frac{|2]\langle q|}{\langle q2\rangle}.
\ee

\subsection{Computation}

We consider the sum of two triangle diagrams. It is enough to consider only one of them, because the other one is obtained by permuting two of the external legs of the first. The diagram we consider is
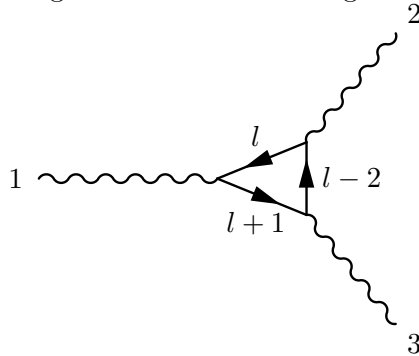
\be
\begin{gathered}
\begin{fmfgraph*}(150,110)
\fmfleft{i1}
\fmfright{o3,o2}
\fmf{photon}{i1,v1}
\fmf{photon}{o2,v2}
\fmf{photon}{o3,v3}
\fmf{fermion,label=$l$,l.side=right}{v2,v1}
\fmf{fermion,label=$l+1$}{v1,v3}
\fmf{fermion,label=$l-2$}{v3,v2}
\fmflabel{1}{i1}
\fmflabel{2}{o2}
\fmflabel{3}{o3}
\end{fmfgraph*}
\end{gathered}\nonumber
\ee

To show how the usual anomaly calculation is reproduced in our notations, we insert the momentum $1$ in the first leg, as well as positive helicity states in legs 2,3. Our Feynman rules are explained in the Appendix, and result in the following Feynman integral
\be 
\label{tri}
\mathcal{A}=\int \frac{d^4l}{(2\pi)^4}\frac{\langle 1|l|2]\langle q|l+1|1]\langle q|l-2|3]}{l^2(l+1)^2(l-2)^2\langle q2\rangle\langle q3\rangle}
\ee
Using (\ref{k1}) the above can be rewritten in terms of a product of momentum spinors as
\be 
\label{tri2}
\mathcal{A}=\int \frac{d^4l}{(2\pi)^4}\frac{\langle q|(l+1)\circ 1\circ l|2]\langle q|l-2|3]}{l^2(l+1)^2(l-2)^2\langle q2\rangle\langle q3\rangle},
\ee
where the notation is that e.g. $1\circ l|2]$ is a primed spinor $1_{M'}{}^{N} l_N{}^{N'} 2_{N'}$. The rule is that the spinor indices are always contracted from top on the left to bottom on the right of an expression. Writing $1=l+1-l$ we expand using 
\be 
\label{l2-ident}
l\circ l = \frac{1}{2} l^2 \id,
\ee 
where $l^2=l\cdot l$ is the norm squared of a vector $l$, and cancel factors of momentum squared from numerator and denominator. We get
\be 
\label{tri3}
\mathcal{A}=\frac{1}{2}\int \frac{d^4l}{(2\pi)^4}\Big[\frac{\langle q|l|2]\langle q|l-2|3]}{l^2(l-2)^2\langle q2\rangle\langle q3\rangle}-\frac{\langle q|l+1|2]\langle q|l-2|3]}{(l-2)^2(l+1)^2\langle q2\rangle\langle q3\rangle}\Big].
\ee
 Thus, we have converted the original linearly divergent loop integral into the difference of two quadratically divergent ones. 

Let us now consider the first integral in (\ref{tri3}). The loop momentum dependent part of this integral is
$$
\int \frac{d^4l}{(2\pi)^4}\frac{l_\mu (l-2)_\nu}{l^2(l-2)^2},
$$
and so it contains a quadratically divergent, as well as linearly divergent part. However, by Lorentz invariance, the linearly divergent part with $l_\mu 2_\nu$ in the numerator can only give a result proportional to $2_\mu 2_\nu$. This vanishes because of $l|2]$ contraction giving $[22]=0$. Thus, there is only the quadratically divergent part
\be\label{ll-int}
\int \frac{d^4l}{(2\pi)^4}\frac{l_\mu l_\nu}{l^2(l-2)^2}.
\ee
However, in any Lorentz invariant regularisation this can only be proportional to $(2)^2 \eta_{\mu\nu}$, which is zero because the momentum $2$ is null, or to $2_\mu 2_\nu$. We thus see that the first integral in (\ref{tri3}) must be a multiple of 
$$ \langle q|2|2]\langle q|2|3] =  \langle q2\rangle [22]\langle q|2|3]=0$$
because the spinor contraction $[22]=0$. 

For the second integral in (\ref{tri3}), we see that shifting the loop momentum $l\rightarrow l+2$ makes it proportional to
$$ \int \frac{d^4l}{(2\pi)^4}\frac{\langle q|l-3|2]\langle q|l|3]}{l^2(l-3)^2}.$$
Again the previous argument applies and we see that the integral vanishes owing to $[33]=0$. Thus, the amplitude in question, if non-zero, is given by the shift. 

Before evaluating the shift we notice that there is an ambiguity for how to deal with the second term in (\ref{tri3}). Indeed, we could instead shift $l\to l-1$. This gives instead
$$ \int \frac{d^4l}{(2\pi)^4}\frac{\langle q|l|2]\langle q|l+3|3]}{l^2(l+3)^2},$$
which is again zero by the same argument. It is clear that in order for the final result to be unambiguous the two different shifts must produce the same result. 

We evaluate the shift by the standard method reviewed in the appendix. Thus, we put
\be 
f(l)=\frac{\langle q|l+1|2]\langle q|l-2|3]}{(l-2)^2(l+1)^2}.
\ee 
We first evaluate the $l\to l+2$ shift. We have, for the linear part of the shift
\be 
\im \lim_{l\to\infty} \int \frac{d\Omega}{(2\pi)^4} 2_\mu l^\mu l^2 f(l).
\ee
We now need to expand the denominator in the large $l$ limit
\be
\frac{1}{(l-2)^2(l+1)^2}= \frac{1}{l^4} \left( 1+ 2 \frac{l\cdot(2-1)}{l^2} + O(\frac{1}{l^2})\right),
\ee
because as we will see below, the subleading term also gives a contribution. 
We then get for the linear part of the shift
\be
\im \lim_{l\to\infty} \int \frac{d\Omega}{(2\pi)^4} 2_\mu l^\mu \frac{\langle q|l+1|2]\langle q|l-2|3]}{l^2} \left( 1+ 2 \frac{l\cdot(2-1)}{l^2} \right).
\ee
Only the quadratic and quartic in $l$ part of the numerator contributes after averaging over the directions. There are two such quadratic parts, with numerators 
$$ 2_\mu l^\mu \langle q|l|2]\langle q|-2|3] \qquad {\rm and}\qquad 2_\mu l^\mu \langle q|1|2]\langle q|l|3].$$
Using
\be\label{shift-ident}
\int \frac{d\Omega}{(2\pi)^4} \frac{l_\mu l_\nu}{l^2} = \frac{2\pi^2}{(2\pi)^4} \frac{1}{4} \eta_{\mu\nu}=\frac{\eta_{\mu\nu}}{32\pi^2},
\ee
where $2\pi^2$ comes as the volume of the unit 3-sphere, we see that the first quadratic part produces a $[22]$ contraction, and is thus zero. The result comes from the second part, and is given by
\be\label{res-1}
\im \lim_{l\to\infty} \int \frac{d\Omega}{(2\pi)^4} 2_\mu l^\mu l^2 f(l)= \frac{\im}{32\pi^2}  \langle q|1|2]\langle q|2|3]= \frac{\im}{32\pi^2}  \langle q1\rangle [12]\langle q2\rangle [23].
\ee
The integral of the quartic in $l$ part is 
\be\label{anom-lin-quartic}
2\im \lim_{l\to\infty} \int \frac{d\Omega}{(2\pi)^4} 2_\mu l^\mu (2-1)_\nu l^\nu \frac{\langle q|l|2]\langle q|l|3]}{l^4}  .
\ee
It is computed using 
\be\label{shift-ident-4l}
\int \frac{d\Omega}{(2\pi)^4} \frac{l_\mu l_\nu l_\rho l_\sigma}{l^4} =\frac{1}{32\cdot 6\pi^2}(\eta_{\mu\nu}\eta_{\rho\sigma} + \eta_{\mu\sigma} \eta_{\nu\rho}+ \eta_{\mu\rho}\eta_{\nu\sigma}).
\ee
The coefficient in this formula is checked by e.g. doing the $\rho\sigma$ contraction, after which it must reproduce (\ref{shift-ident}). There is only one possible contraction in (\ref{anom-lin-quartic}), which gives 
\be\label{res-2}
\frac{2\im }{32\cdot 6\pi^2} \langle q|2-1|2]\langle q|2|3]= - \frac{\im }{3\cdot 32\pi^2} \langle q|1|2]\langle q|2|3].
\ee
Adding (\ref{res-1}) and (\ref{res-2}) gives
\be\label{anom-result}
\frac{2\im}{3\cdot 32\pi^2}  \langle q|1|2]\langle q|2|3]
\ee
for the linear part of the shift.

And for the quadratic part of the shift, the integral in 
\be\label{anom-quadr-shift}
\lim_{l\to\infty} \int \frac{d\Omega}{(2\pi)^4}  2_\mu 2_\nu l^\mu l^2\frac{\partial}{\partial l_\nu}\frac{\langle q|l+1|2]\langle q|l-2|3]}{(l-2)^2(l+1)^2}
\ee
is easily seen to be zero. Indeed, when the derivative hits the denominator one produces a multiple of 
\be
\lim_{l\to\infty} \int \frac{d\Omega}{(2\pi)^4}  2_\mu 2_\nu l^\mu l^\nu \frac{\langle q|l+1|2]\langle q|l-2|3]}{l^4}.
\ee
One must keep only the quartic in $l$ term in the numerator, but then the averaging over the directions of $l^\mu$ produces factors of the metric tensor that cause one of the copies of $2_\mu$ contract with the spinor $|2]$, which gives zero, or the two copies of $2_\mu$ contract, which again gives zero. So, the derivative in (\ref{anom-quadr-shift}) must hit the numerator. It cannot hit the quantity $\langle q|l+1|2]$ because then there is again a factor of $\langle q| 2|2]\sim [22]=0$. So, it must hit $ \langle q|l-2|3]$. We then have for (\ref{anom-quadr-shift})
\be
\lim_{l\to\infty} \int \frac{d\Omega}{(2\pi)^4}  2_\mu  l^\mu \frac{\langle q|l+1|2]\langle q|2|3]}{l^2},
\ee
and again the averaging causes this to be proportional to $\langle q| 2|2]\sim [22]=0$. So, the quadratic part of the shift is zero.

To check the consistency of our scheme, let us calculate the shift $l\to l-1$. The linear part of the shift is
\be 
-\im \lim_{l\to\infty} \int \frac{d\Omega}{(2\pi)^4} 1_\mu l^\mu l^2 f(l)=-\im \lim_{l\to\infty} \int \frac{d\Omega}{(2\pi)^4} 1_\mu l^\mu \frac{\langle q|l+1|2]\langle q|l-2|3]}{l^2}\left( 1+ 2 \frac{l\cdot(2-1)}{l^2} \right).
\ee
Again, the quadratic part of the numerator consists of two terms
$$ 1_\mu l^\mu \langle q|l|2]\langle q|-2|3] \qquad {\rm and}\qquad  1_\mu l^\mu \langle q|1|2]\langle q|l|3] .$$
Now both terms give a non-zero answer after integration, and we have
$$
-\im \lim_{l\to\infty} \int \frac{d\Omega}{(2\pi)^4} 1_\mu l^\mu l^2 f(l)=- \frac{\im}{32\pi^2} (\langle q|1|2]\langle q|-2|3]+  \langle q|1|2]\langle q|1|3]).$$
Replacing $1=-2-3$ in the last term and using $[33]=0$ we see that that the terms in brackets are the same and so 
\be\label{anom-altern-lin-1}
-\im \lim_{l\to\infty} \int \frac{d\Omega}{(2\pi)^4} 1_\mu l^\mu l^2 f(l)= \frac{2\im}{32\pi^2} \langle q|1|2]\langle q|2|3].
\ee
The quartic part of the integrand is
\be
-2\im \lim_{l\to\infty} \int \frac{d\Omega}{(2\pi)^4} 1_\mu l^\mu (2-1)_\nu l^\nu \frac{\langle q|l|2]\langle q|l|3]}{l^4}.
\ee
Using (\ref{shift-ident-4l}) this gives two terms
\be
-\frac{2\im}{32\cdot 6\pi^2} (\langle q|1|2]\langle q|2-1|3] + \langle q|2-1|2]\langle q|1|3]).
\ee
Using $-1=2+3$ in the first term and $1=-2-3$ we get a factor of 3 in the brackets and this becomes
\be\label{anom-altern-lin-2}
-\frac{\im}{32\pi^2} \langle q|1|2]\langle q|2|3].
\ee
Thus, collecting (\ref{anom-altern-lin-1}) and (\ref{anom-altern-lin-2}) we get $(\im/32\pi^2) \langle q|1|2]\langle q|2|3]$. 

Let us also compute the quadratic part of the shift, which is now non-vanishing. It is equal to
\be\label{anom-altern}
\frac{\im}{2} \lim_{l\to\infty} \int \frac{d\Omega}{(2\pi)^4}  1_\mu 1_\nu l^\mu l^2\frac{\partial}{\partial l_\nu}\frac{\langle q|l+1|2]\langle q|l-2|3]}{(l-2)^2(l+1)^2}.
\ee
There are many non-zero terms in this case. First, when the derivative hits the denominator we get, in the large $l$ limit
\be\label{anom-den-hit}
\frac{\im}{2} (-4) \lim_{l\to\infty} \int \frac{d\Omega}{(2\pi)^4}  1_\mu 1_\nu l^\mu l^\nu  \frac{\langle q|l+1|2]\langle q|l-2|3]}{l^4}.
\ee
Using (\ref{shift-ident-4l}) we get
\be\label{anom-den-hit-res}
-\frac{2\im}{3\cdot 32\pi^2} \langle q|1|2]\langle q|1|3] = \frac{2\im}{3\cdot 32\pi^2} \langle q|1|2]\langle q|2|3].
\ee
There are also two terms when the derivative in (\ref{anom-altern}) hits the numerator. These are, again in the large $l$ limit
\be
\frac{\im}{2} \lim_{l\to\infty} \int \frac{d\Omega}{(2\pi)^4}  1_\mu  l^\mu \frac{\langle q|1|2]\langle q|l-2|3]+\langle q|l+1|2]\langle q|1|3] }{l^2}.
\ee
Using (\ref{shift-ident}) we get for these terms
\be\label{anom-altern-num-hit-res}
\frac{\im}{32\pi^2} \langle q|1|2]\langle q|1|3]= - \frac{\im}{32\pi^2} \langle q|1|2]\langle q|2|3].
\ee
This exactly cancels what comes from the linear part of the shift, and so the result of the $l\to l-1$ shift equals to (\ref{anom-den-hit-res}), which matches (\ref{anom-result}). So, the rules we use are consistent. 

\bigskip
\noindent{\bf Final result.}
Overall, the diagram (\ref{tri}) is given by (minus) the shift. The minus sign comes because the diagram with the integrand $f(l+2)$ is zero. The calculated shift is $f(l+2)-f(l)$, and so the original diagram is minus the shift. The shift was computed to be (\ref{anom-result}), and there is an extra factor of minus a half in (\ref{tri3}). This gives for the following answer for the triangle diagram
\be
\frac{i}{96\pi^2}\frac{[23][12]\langle q1\rangle}{\langle q3\rangle}.
\ee 
Adding to this the second diagram that can be obtained as ($3\leftrightarrow 2$), we have the final result of the triangle anomaly:
\be 
\label{shiftA3}
\frac{i}{96\pi^2}\Big(\frac{[23][12]\langle q1\rangle}{\langle q3\rangle}+\frac{[32][13]\langle q1\rangle}{\langle q2\rangle}\Big).
\ee
This can be simplified using the momentum conservation 
$$ |1]\langle 1| +  |2]\langle 2|+ |3]\langle 3|=0.$$
Then, multiplying with e.g. $[2|$ and $|q\rangle$ we get $[21]\langle 1q\rangle + [23]\langle 3q\rangle=0$, from which $\langle q1\rangle/\langle q3\rangle = - [23]/[21]$. Similarly
$\langle q1\rangle/\langle q2\rangle = - [32]/[31]$. This means that both terms in (\ref{shiftA3}) are equal and so we get for the anomaly
\be\label{trangle-anomaly}
\frac{i}{48\pi^2}[23]^2,
\ee
which is the expected result. Indeed, the anomaly is a multiple of $\epsilon^{\mu\nu\rho\sigma}F_{\mu\nu} F_{\rho\sigma}$. For the polarisations used only the anti-self-dual (ASD) part of the field strength is different from zero. In spinor notations this is $F_{MM'NN'}\sim F_{M'N'}\epsilon_{MN}$. Moreover, for a polarisation corresponding to a null momentum $k_M{}^{M'}=k_M k^{M'}$, the ASD part of the field strength is $F_{M'N'}\sim k_{M'}k_{N'}$. Thus, (\ref{trangle-anomaly}) is the expected result. 

\section{Box Amplitude}
\label{sec:box}

\vspace{1em}
\vspace{1em}
\vspace{1em}
\vspace{1em}
\be
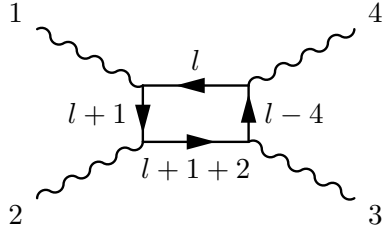
\begin{fmfgraph*}(120,80)
\fmftop{i1,i4}
\fmfbottom{o2,o3}
\fmf{photon}{i1,v1}
\fmf{photon}{i4,v4}
\fmf{photon}{o2,v2}
\fmf{photon}{o3,v3}
\fmf{fermion,label=$l$,l.side=right}{v4,v1}
\fmf{fermion,label=$l+1$,l.side=right}{v1,v2}
\fmf{fermion,label=$l+1+2$}{v2,v3}
\fmf{fermion,label=$l-4$}{v3,v4}
\fmflabel{1}{i1}
\fmflabel{4}{i4}
\fmflabel{2}{o2}
\fmflabel{3}{o3}
\end{fmfgraph*}\nonumber
\ee

Let us consider one of the box diagrams, the one shown in the figure. In index-free notations, this diagram with positive helicity states inserted into all the legs is given by
\be \label{boxamp}
i\mathcal{A}=\int \frac{d^4l}{(2\pi)^4}\frac{\langle q|l|4]\langle q|l+1|1]\langle q|l+1+2|2]\langle q|l-4|3]}{l^2(l+1)^2(l+1+2)^2(l-4)^2\prod_{j=1}^4\langle qj\rangle}
\ee
\\~\\
We start by multiplying the numerator and denominator of the integral by $\langle 43\rangle$, and using $\langle q|l-4|3]=[3|l-4|q\rangle$, which is confirmed by raising-lowering a pair of spinor indices. This allows to write the amplitude as
\be 
\label{amp}
i\mathcal{A}=\int \frac{d^4l}{(2\pi)^4}\frac{\langle q| l \circ 4 \circ 3\circ (l-4)|q\rangle\langle q|l+1|1]\langle q|l+1+2|2]}{l^2(l+1)^2(l+1+2)^2(l-4)^2\prod_{j=1}^4\langle qj\rangle\langle 43\rangle}
\ee
We then replace $4=l-(l-4)$, and use the identity (\ref{l2-ident}), which gives
\be 
\label{amp12}
i\mathcal{A}=\frac{1}{2}\int \frac{d^4l}{(2\pi)^4}\frac{\langle q|3\circ (l-4)|q\rangle\langle q|l+1|1]\langle q|l+1+2|2]}{(l+1)^2(l+1+2)^2(l-4)^2\prod_{j=1}^4\langle qj\rangle\langle 43\rangle}\\\nonumber-\int \frac{d^4l}{(2\pi)^4}\frac{\langle q| l \circ (l-4)\circ 3\circ (l-4)|q\rangle\langle q|l+1|1]\langle q|l+1+2|2]}{l^2(l+1)^2(l+1+2)^2(l-4)^2\prod_{j=1}^4\langle qj\rangle\langle 43\rangle}
\ee
Next we use the spinor identity
\be
\label{id}
A\circ B=-B\circ A+(A\cdot B) \mathds{1},
\ee
where $(A\cdot B)$ is the metric pairing, which is true for arbitrary for any arbitrary rank 2 mixed spinors $A$ and $B$. Using this identity we have
\be\nonumber
(l-4)\circ 3\circ (l-4) = - (l-4)\circ (l-4)\circ 3 + (l-4) ((l-4)\cdot 3) \\ \nonumber
= - \frac{1}{2}(l-4)^2 3+ ((l-4)\cdot 3) (l-4).
\ee
On the other hand, using $3^2=0$, we can write
$$((l-4)\cdot 3) = -\frac{1}{2}( (l-4-3)^2 - (l-4)^2).$$
This gives
\be\label{ABA}
(l-4)\circ 3\circ (l-4) = \frac{1}{2} (l-4)^2 (l-4-3) - \frac{1}{2} (l-4-3)^2 (l-4). 
\ee
Replacing $l-4-3=l+1+2$, and cancelling some of the denominators we get
\be 
2i\mathcal{A}=\int \frac{d^4l}{(2\pi)^4}\frac{\langle q|3\circ(l-4)|q\rangle\langle q|l+1|1]\langle q|l+1+2|2]}{(l+1)^2(l+1+2)^2(l-4)^2\prod_{j=1}^4\langle qj\rangle\langle 43\rangle}\\\nonumber+\int \frac{d^4l}{(2\pi)^4}\frac{\langle q| l \circ (l-4)|q\rangle\langle q|l+1|1]\langle q|l+1+2|2]}{l^2(l+1)^2(l-4)^2\prod_{j=1}^4\langle qj\rangle\langle 43\rangle}\\\nonumber-\int \frac{d^4l}{(2\pi)^4}\frac{\langle q| l \circ (l-3-4)|q\rangle\langle q|l+1|1]\langle q|l+1+2|2]}{l^2(l+1)^2(l+1+2)^2\prod_{j=1}^4\langle qj\rangle\langle 43\rangle}.
\ee
In the first of these integrals the numerator is cubic in $l$, and so it is linearly divergent. The second and third integrals have four powers of $l$ in the numerator, and so are nominally quadratically divergent. However, one of the factors of $l$ can be eliminated immediately. Indeed, the expression $\langle q| l | l | q\rangle =(1/2) l^2 \langle qq\rangle=0$, and so 
$\langle q| l |l-4| q\rangle= \langle q| l |-4 | q\rangle$, and similarly in the last integral. Thus,
\be
2i\mathcal{A}=2i\mathcal{A}_1+2i\mathcal{A}_2+2i\mathcal{A}_3,
\ee
with
\be\label{amp3}
2i\mathcal{A}_1=\int \frac{d^4l}{(2\pi)^4}\frac{\langle q|3\circ (l-4)|q\rangle\langle q|l+1|1]\langle q|l+1+2|2]}{(l+1)^2(l+1+2)^2(l-4)^2\prod_{j=1}^4\langle qj\rangle\langle 43\rangle}\\\nonumber
2i\mathcal{A}_2=-\int \frac{d^4l}{(2\pi)^4}\frac{\langle q|l \circ 4|q\rangle\langle q|l+1|1]\langle q|l+1+2|2]}{l^2(l+1)^2(l-4)^2\prod_{j=1}^4\langle qj\rangle\langle 43\rangle}\\\nonumber
2i\mathcal{A}_3=\int \frac{d^4l}{(2\pi)^4}\frac{\langle q|l \circ (3+4)|q\rangle\langle q|l+1|1]\langle q|l+1+2|2]}{l^2(l+1)^2(l+1+2)^2\prod_{j=1}^4\langle qj\rangle\langle 43\rangle},
\ee
with all integrals linearly divergent, and of the type previously considered in the anomaly calculation. The idea of the calculation now is to consider these integrals one by one, and convert them to quadratically divergent integrals as we have done in the anomaly case. These quadratically divergent integrals are then reduced to shifts, which are evaluated by the standard methods. 

\subsection*{Calculation of $\mathcal{A}_3$}

Let us first consider the last integral in (\ref{amp3}). We rewrite $\langle q| l+1+2| 2]=[2|l+1+2| q\rangle$, and thus have
$$\langle q|l+1|1]\langle q|l+1+2|2]\langle 12\rangle = \langle q| (l+1)\circ   1 \circ  2 \circ  (l+1+2) |q\rangle.$$
We then replace $2=l+1+2-(l+1)$ to get
$$\langle q| (l+1)\circ   1 \circ  2 \circ  (l+1+2) |q\rangle= \frac{1}{2} \langle q| (l+1)\circ 1 | q\rangle (l+1+2)^2 - \langle q| (l+1) \circ 1 \circ (l+1) \circ (l+1+2) |q\rangle.$$
We then use the analog of (\ref{ABA}) 
$$(l+1)\circ 1\circ (l+1) = \frac{1}{2} (l+1)^2 l - \frac{1}{2} l^2 (l+1).$$
Overall, we get
\be\nonumber
2\langle q| (l+1)\circ   1 \circ  2 \circ  (l+1+2) |q\rangle = \langle q| (l+1)\circ 1 | q\rangle (l+1+2)^2 - (l+1)^2 \langle q| l \circ (l+1+2)| q\rangle \\ \nonumber
+ l^2 \langle q| (l+1) \circ (l+1+2)| q\rangle.
\ee
The last two terms can be further simplified by removing one of the factors of the loop momentum, as causing the $\langle qq\rangle$ contraction
$$2\langle q| (l+1)\circ   1 \circ  2 \circ  (l+1+2) |q\rangle  = \langle q| (l+1)\circ  1 | q\rangle (l+1+2)^2 - (l+1)^2 \langle q| l \circ  (1+2)| q\rangle + l^2 \langle q| (l+1) \circ 2| q\rangle.$$
Thus, we get 
\be\label{M3}
4\im{\mathcal A}_3 = \frac{1}{\prod_{j=1}^4\langle qj\rangle\langle 12\rangle\langle 43\rangle }\int \frac{d^4l}{(2\pi)^4} \Big[ \frac{\langle q|l\circ (3+4)|q\rangle\langle q| (l+1)\circ 1 | q\rangle }{l^2(l+1)^2}  \\ \nonumber - \frac{\langle q|l\circ (3+4)|q\rangle\langle q| l \circ (1+2) | q\rangle }{l^2(l+1+2)^2}+\frac{\langle q|l \circ (3+4)|q\rangle\langle q| (l+1) \circ 2| q\rangle }{(l+1)^2(l+1+2)^2} \Big].
\ee
We now use the same argument as in (\ref{ll-int}) to argue that some terms are zero. Indeed the first integral with $l_\mu l_\nu$ in the numerator can only give $\eta_{\mu\nu}$, which causes $q$'s to contract, or $1_\mu 1_\nu$, which produces $1^2=0$. In the second term, similarly, $l_\mu l_\nu$ in the numerator can only give $(1+2)_\mu (1+2)_\nu$. This gives a $\langle qq\rangle=0$ contraction. 

In the third term, we can argue that the integral is zero after a shift. The easiest shift that does the job is $l\to l-1-2$. This converts the integral to
$$ \int \frac{d^4l}{(2\pi)^4} \frac{\langle q|(l-1-2)\circ(3+4)|q\rangle\langle q| (l-2) \circ 2| q\rangle }{l^2(l-2)^2}.$$
Using $-1-2=3+4$ and (\ref{l2-ident}) we can see that this equals
$$ \int \frac{d^4l}{(2\pi)^4} \frac{\langle q|l \circ(3+4)|q\rangle\langle q| l \circ 2| q\rangle }{l^2(l-2)^2}.$$
Again the same argument as before shows that the integral is proportional to $2^2=0$. Thus, the part ${\mathcal M}_3$ reduces to a shift. 

We now compute the shift. The linear part of the shift is given by
$$ -\im \lim_{l\to \infty} \int \frac{d\Omega}{(2\pi)^4} (1+2)_\mu l^\mu  \frac{\langle q|l\circ (3+4)|q\rangle\langle q| (l+1) \circ  2| q\rangle }{l^2}\left( 1-2 \frac{2(l\cdot 1)+(l\cdot 2)}{l^2}\right),$$
where we again kept the subleading term. Using (\ref{shift-ident}) we see that there is no contribution coming from the quadratic in $l$ part of the numerator. Using (\ref{shift-ident-4l}) we get the contribution from the quartic in $l$ part
\be\label{M3-linear}
  -\frac{\im}{3\cdot 32\pi^2} \langle q| 1 \circ 2| q\rangle^2.
\ee
The quadratic part of the shift is
\be
\frac{\im}{2} \lim_{l\to \infty} \int \frac{d\Omega}{(2\pi)^4} (1+2)_\mu l^\mu  l^2 (1+2)_\nu \frac{\partial}{\partial l_\nu} \frac{\langle q|l\circ (3+4)|q\rangle\langle q| (l+1) \circ  2| q\rangle }{(l+1)^2(l+1+2)^2}.
\ee
When the derivative hits the denominator we get, in the large $l$ limit
\be
-2\im \lim_{l\to \infty} \int \frac{d\Omega}{(2\pi)^4} (1+2)_\mu l^\mu  (1+2)_\nu l^\nu \frac{\langle q|l\circ (3+4)|q\rangle\langle q| l \circ  2| q\rangle }{l^4},
\ee
which, using (\ref{shift-ident-4l}),  gives zero. 
When the derivative hits the numerator it must hit the second term, but then the integration causes $(1+2)\circ(3+4)\sim \id$ contraction, which causes $\langle qq\rangle=0$. Thus, there is no quadratic part of the shift, and the total result is given by (\ref{M3-linear}).

We can check this result by noticing that another shift, namely $l\to l-1$, also gives zero. This converts the integral to
$$ \int \frac{d^4l}{(2\pi)^4} \frac{\langle q|(l-1)\circ(3+4)|q\rangle\langle q| l \circ 2| q\rangle }{l^2(l+2)^2}.$$
Again the same argument as before shows that the integral is proportional to $2^2=0$. Thus, the part ${\mathcal M}_3$ reduces to a shift. 

We now compute the shift. The linear part of the shift is given by
$$ -\im \lim_{l\to \infty} \int \frac{d\Omega}{(2\pi)^4} 1_\mu l^\mu  \frac{\langle q|l\circ (3+4)|q\rangle\langle q| (l+1) \circ  2| q\rangle }{l^2}\left( 1-2 \frac{2(l\cdot 1)+(l\cdot 2)}{l^2}\right),$$
where we again kept the subleading term. Using (\ref{shift-ident}) and (\ref{shift-ident-4l}) we get all terms to cancel, and hence there is no linear part of the shift.

The quadratic part of the shift is
\be
\frac{\im}{2} \lim_{l\to \infty} \int \frac{d\Omega}{(2\pi)^4} 1_\mu l^\mu  l^2 1_\nu \frac{\partial}{\partial l_\nu} \frac{\langle q|l\circ (3+4)|q\rangle\langle q| (l+1) \circ  2| q\rangle }{(l+1)^2(l+1+2)^2}.
\ee
When the derivative hits the denominator we get, in the large $l$ limit
\be
-2\im \lim_{l\to \infty} \int \frac{d\Omega}{(2\pi)^4} 1_\mu l^\mu  1_\nu l^\nu \frac{\langle q|l\circ (3+4)|q\rangle\langle q| l \circ  2| q\rangle }{l^4},
\ee
which gives $(2\im/3\cdot 32\pi^2)\langle q| 1 \circ  2| q\rangle^2$. When the derivative hits the numerator we get two equal contributions, totalling
$-(\im/32\pi^2)\langle q| 1 \circ  2| q\rangle^2$. Adding everything up we get the same result (\ref{M3-linear}) for the shift, thus checking that everything is consistent. 

The result for ${\cal A}_3$ is given by minus the shift. There are also prefactors in (\ref{M3}). Taking all into account gives
\be
{\mathcal A}_3 =  \frac{1}{12\cdot 32\pi^2} \frac{\langle q1\rangle \langle q2\rangle}{\langle q3\rangle \langle q4\rangle } \frac{[12]^2}{\langle 12\rangle\langle 43\rangle }.
\ee

\subsection*{Calculation of $\mathcal{A}_1$}

Next consider the first integral of (\ref{amp3}). We rewrite $\langle q|3\circ(l-4)|q\rangle = \langle q3\rangle [3|l-4|q\rangle $, and then 
multiply the numerator and denominator by $\langle 23\rangle$. This yields
\be 
\label{amp13}
2i\mathcal{A}_{1}=\int \frac{d^4l}{(2\pi)^4}\frac{\langle q|l+1|1]\langle q|(l+1+2)\circ 2\circ 3\circ (l-4)|q\rangle}{(l+1)^2(l+1+2)^2(l-4)^2\langle 23\rangle\langle 43\rangle\langle q1\rangle\langle q2\rangle\langle q4\rangle}
\ee
Then using $3=-(l+1+2)+(l-4)$ and (\ref{l2-ident}) we get 
\be 
\label{amp11}
2i\mathcal{A}_{1}=-\int \frac{d^4l}{(2\pi)^4}\frac{\langle q| l+1|1]\langle q|(l+1+2)\circ 2\circ (l+1+2) \circ(l-4)|q\rangle}{(l+1)^2(l+1+2)^2(l-4)^2\langle 23\rangle\langle 43\rangle\langle q1\rangle\langle q2\rangle\langle q4\rangle}\\+\frac{1}{2}\nonumber\int \frac{d^4l}{(2\pi)^4}\frac{\langle q|l+1|1]\langle q|(l+1+2)\circ 2|q\rangle}{(l+1)^2(l+1+2)^2\langle 32\rangle\langle 43\rangle\langle q1\rangle\langle q2\rangle\langle q4\rangle}
\ee
We then use the analog of (\ref{ABA}), which reads
$$ (l+1+2)\circ 2 \circ(l+1+2) = \frac{1}{2} (l+1+2)^2 (l+1) - \frac{1}{2} (l+1)^2 (l+1+2).$$ 
This gives
\be 
\label{amp111}
4i\mathcal{A}_{1}=-\int \frac{d^4l}{(2\pi)^4}\frac{\langle q| l+1 |1]\langle q|(l+1)\circ (l-4)|q\rangle}{(l+1)^2(l-4)^2\langle 23\rangle\langle 43\rangle\langle q1\rangle\langle q2\rangle\langle q4\rangle}\\\nonumber
+\int \frac{d^4l}{(2\pi)^4}\frac{\langle q| l+1 |1]\langle q|(l+1+2)\circ(l-4)|q\rangle}{(l+1+2)^2(l-4)^2\langle 23\rangle\langle 43\rangle\langle q1\rangle\langle q2\rangle\langle q4\rangle}
\\+\nonumber\int \frac{d^4l}{(2\pi)^4}\frac{\langle q| l+1 |1]\langle q|(l+1+2)\circ 2|q\rangle}{(l+1)^2(l+1+2)^2\langle 23\rangle\langle 43\rangle\langle q1\rangle\langle q2\rangle\langle q4\rangle}
\ee
All these integrals are seen to be zero after a shift, and so their calculation reduces to computation of the shifts.

Let us consider the first integral of (\ref{amp111}). We see that shifting $l\rightarrow l-1$ results into 
\be 
\int \frac{d^4l}{(2\pi)^4}\frac{\langle q| l|1]\langle q|l\circ (l-1-4)|q\rangle}{(l-1-4)^2l^2}=-\int \frac{d^4l}{(2\pi)^4}\frac{\langle q|l|1]\langle q| l\circ(1+4)|q\rangle}{(l-1-4)^2l^2}
\ee
This integral can only depend on $1+4$, and hence vanishes as giving a $\langle qq\rangle$ contraction. 

Let us compute the result of this shift. The integrand is
$$
\frac{\langle q| l+1 |1]\langle q|(l+1)\circ (l-4)|q\rangle}{(l+1)^2(l-4)^2}=\frac{\langle q| l |1]\langle q|(l+1)\circ (2+3)|q\rangle}{(l+1)^2(l-4)^2}.
$$
The linear part of the shift is
\be
-\im \lim_{l\to \infty} \int \frac{d\Omega}{(2\pi)^4} 1_\mu l^\mu  \frac{\langle q| l |1]\langle q|(l+1)\circ (2+3)|q\rangle}{l^2} \left( 1+2 \frac{ (l\cdot 4)-(l\cdot 1)}{l^2}\right).
\ee
The quadratic in $l$ part gives zero, while the quartic part gives $-(\im/3\cdot 32\pi^2) \langle q| 4|1] \langle q| 4\circ 1|q\rangle$. The quadratic part of the shift can be seen to be zero.

Then, recalling that the integral is minus the shift, gives the following answer for the first integral in (\ref{amp111})
\be
-\frac{\im}{3\cdot 32\pi^2} \frac{\langle q| 4|1] \langle q| 4\circ 1|q\rangle  }{\langle 23\rangle\langle 43\rangle\langle q1\rangle\langle q2\rangle\langle q4\rangle}.
\ee

Let us consider the second integral of (\ref{amp111}). We see that shifting $l\rightarrow l+4$ results into 
\be 
\label{amppart1}
\int \frac{d^4l}{(2\pi)^4}\frac{\langle q| l+4 |1]\langle q|(l-3)\circ l |q\rangle}{(l-3)^2l^2}=-\int \frac{d^4l}{(2\pi)^4}\frac{\langle q|l+4|1]\langle q|3\circ l|q\rangle}{(l-3)^2l^2}
\ee
\\
There is a quadratically divergent part here that vanishes as giving a $[33]$ contraction, as well as a linearly divergent part that again vanishes for the same reason. So, after the shift the result is zero. 

Let us compute the shift. The relevant integrand is
\be 
\frac{\langle q|l+1|1]\langle q|(l+1+2)\circ(l-4)|q\rangle}{(l+1+2)^2(l-4)^2}=-\frac{\langle q|l|1]\langle q|3\circ(l-4)|q\rangle}{(l+1+2)^2(l-4)^2},
\ee 
and then the linear part of the shift is
\be
-\im \lim_{l\to \infty} \int \frac{d\Omega}{(2\pi)^4} 4_\mu l^\mu \frac{\langle q|l|1]\langle q|3\circ(l-4)|q\rangle}{l^2}\left( 1+2 \frac{ 2(l\cdot 4)+(l\cdot 3)}{l^2}\right).
\ee
The quadratic part in $l$ gives $(\im/32\pi^2)  \langle q| 4|1] \langle q|3\circ 4|q \rangle$. The quartic part gives $$-(\im/3\cdot 32\pi^2)( 4 \langle q| 4|1] \langle q|3\circ 4|q \rangle+  \langle q| 3|1] \langle q|3\circ 4|q \rangle.$$ Adding these two contributions gives
$$-\frac{\im}{3\cdot 32\pi^2} \langle q| 4+3|1] \langle q|3\circ 4|q \rangle $$

The quadratic part of the shift is 
\be 
-\frac{\im}{2} \lim_{l\to\infty}  \int \frac{d\Omega}{(2\pi)^4} 4_{\mu}4_{\nu}l^{\mu}l^2\frac{\partial}{\partial{l_{\nu}}}\frac{\langle q|l|1]\langle q|3\circ (l-4)|q\rangle}{(l+1+2)^2(l-4)^2}
\ee
When the derivative hits the denominator we get
\be 
2\im \lim_{l\to\infty}  \int \frac{d\Omega}{(2\pi)^4} 4_{\mu}4_{\nu}l^{\mu} l^\nu \frac{\langle q|l|1]\langle q|3\circ l|q\rangle}{l^4}= \frac{2\im}{3\cdot 32\pi^2} \langle q|4|1]\langle q|3\circ 4|q\rangle.
\ee
When the derivative hits the numerator we get two equal contributions resulting in $$-\frac{\im}{32\pi^2} \langle q|4|1]\langle q|3\circ 4|q\rangle.$$ Adding the two contributions we get for the quadratic part of the shift
$$- \frac{\im}{3\cdot 32\pi^2} \langle q| 4 |1] \langle q|3\circ 4|q \rangle .$$
Overall, we get for this shift
$$-\frac{\im}{3\cdot 32\pi^2} (2 \langle q| 4|1] +\langle q|3|1]) \langle q|3\circ 4|q \rangle. $$
We have checked this result by computing the $l\to l-1-2$ shift instead. This can be further simplified using the momentum conservation. Writing $2k_4+k_3=k_4-(k_1+k_2)$ shows that we can rewrite this result as
$$-\frac{\im}{3\cdot 32\pi^2} \langle q| 4-2|1] \langle q|3\circ 4|q \rangle. $$

The integral is minus the shift. Thus, we get the following answer for the second term
\be
\frac{\im}{3\cdot 32\pi^2} \frac{\langle q| 4-2|1] \langle q|3\circ 4|q \rangle} {\langle 23\rangle \langle 43\rangle \langle q1\rangle \langle q2\rangle \langle q4\rangle}.
\ee

Finally, we consider the last term in (\ref{amp111}). The simplest shift that does the job is again $l\to l-1$. The relevant integrand is then
$$ \frac{\langle q| l|1]\langle q|(l+1)\circ 2|q\rangle}{(l+1)^2(l+1+2)^2}.$$
The linear part of the shift is
\be
-\im \lim_{l\to \infty} \int \frac{d\Omega}{(2\pi)^4} 1_\mu l^\mu \frac{\langle q|l|1]\langle q|(l+1)\circ 2|q\rangle}{l^2}\left( 1-2 \frac{ 2(l\cdot 1)+(l\cdot 2)}{l^2}\right).
\ee
The quadratic part gives no contribution, while the quartic part gives $ (\im/3\cdot 32\pi^2) \langle q|2|1] \langle q|1\circ 2|q\rangle$. There is no contribution from the quadratic part of the shift. 
The full answer for the last term in (\ref{amp111}) is then
\be
-\frac{\im}{3\cdot 32\pi^2} \frac{ \langle q|2|1] \langle q|1\circ 2|q\rangle }{\langle 23\rangle \langle 43\rangle  \langle q1\rangle \langle q2\rangle \langle q4\rangle}.
\ee

Adding up all 3 contributions we get
\be
\mathcal{A}_1= -\frac{1}{12\cdot 32\pi^2}\frac{\langle q| 4|1] \langle q| 4\circ 1|q\rangle + \langle q| 2-4|1] \langle q|3\circ 4|q \rangle+ \langle q|2|1] \langle q|1\circ 2|q\rangle }{\langle 23\rangle \langle 43\rangle  \langle q1\rangle \langle q2\rangle \langle q4\rangle}.
\ee

\subsection*{Simplification}

The result for $\mathcal{A}_1$ is quite long, and it is desirable to simplify it if possible before any other manipulations. To this end, we multiply the numerator and denominator with $\langle 1 q\rangle$, which converts the numerator into
$$\langle q| 4\circ 1 |q\rangle  \langle q| 4\circ 1|q\rangle + \langle q| (2-4)\circ 1 |q\rangle \langle q|3\circ 4|q \rangle+ \langle q|2\circ 1 |q\rangle \langle q|1\circ 2|q\rangle .$$
This can be simplified using the momentum conservation. Thus, we notice that there are relations of the type
$$ 1\circ 2 + 1\circ 3+1\circ 4 = - 4\circ 4 = - \frac{1}{2} 4^2 \id =0.$$
There are 4 such relations among 6 different momentum products. This makes it possible to choose any 3 of them as independent, and write the other 3 in terms of the basis chosen. Let us choose as independent the products
$$ 1\circ 2, \quad 1\circ 4, \quad 2\circ 3.$$
This choice is motivated by the fact that $1\circ 2, 1\circ 4$ appears in $\mathcal{M}_1$. 
We then get
$$ 3\circ 4 = - 1\circ 2- 1\circ 4 + 2\circ 3.$$
Substituting this into the numerator of the amplitude we notice that there are multiple cancellations, and the result is
\be\label{res-m1}
\mathcal{A}_1= \frac{1}{12\cdot 32\pi^2}\frac{\langle q|1\circ(4-2) |q \rangle \langle q|2\circ 3 | q \rangle } {\langle 23\rangle \langle 43\rangle  \langle q1\rangle^2 \langle q2\rangle \langle q4\rangle}.
\ee

\subsection*{Calculation of $\mathcal{A}_2$}

Next we consider $\mathcal{A}_2$. Writing $\langle q|l\circ 4|q\rangle = \langle q| l|4] \langle 4q\rangle$, and multiplying the numerator and denominator by $\langle 41\rangle$ yields
\be 
\label{amp23}
2i\mathcal{A}_{2}=\int \frac{d^4l}{(2\pi)^4}\frac{\langle q|l+1+2|2]\langle q|l \circ 4 \circ 1\circ (l+1)|q\rangle}{l^2(l+1)^2(l-4)^2\langle 41\rangle\langle 43\rangle\langle q1\rangle\langle q2\rangle\langle q3\rangle}
\ee
Using $1=l+1-l$ and (\ref{l2-ident}) we get
\be 
\label{amp231}
2i\mathcal{A}_{2}=\frac{1}{2}\nonumber\int \frac{d^4l}{(2\pi)^4}\frac{\langle q|l+1|2]\langle q|l \circ 4|q\rangle}{l^2(l-4)^2\langle 41\rangle\langle 43\rangle\langle q1\rangle\langle q2\rangle\langle q3\rangle} \\ \nonumber
-\int \frac{d^4l}{(2\pi)^4}\frac{\langle q|l+1|2]\langle q| l \circ 4\circ l\circ (l+1)|q\rangle}{l^2(l+1)^2(l-4)^2\langle 41\rangle\langle 43\rangle\langle q1\rangle\langle q2\rangle\langle q3\rangle}
\ee
The second integral is again manipulated using an analog of (\ref{ABA})
$$l\circ 4\circ l = \frac{1}{2} l^2 (l-4) - \frac{1}{2} (l-4)^2 l.$$
This gives
\be 
\label{amp232}
4i\mathcal{A}_{2}=\nonumber\int \frac{d^4l}{(2\pi)^4}\frac{\langle q|l+1|2]\langle q|l\circ 4|q\rangle}{l^2(l-4)^2\langle 41\rangle\langle 43\rangle\langle q1\rangle\langle q2\rangle\langle q3\rangle} \\ \nonumber
- \int \frac{d^4l}{(2\pi)^4}\frac{\langle q|l+1|2]\langle q| (l-4) \circ (l+1) |q\rangle}{(l+1)^2(l-4)^2\langle 41\rangle\langle 43\rangle\langle q1\rangle\langle q2\rangle\langle q3\rangle}\\ \nonumber
+\int \frac{d^4l}{(2\pi)^4}\frac{\langle q|l+1|2]\langle q| l \circ 1 |q\rangle}{l^2(l+1)^2\langle 41\rangle\langle 43\rangle\langle q1\rangle\langle q2\rangle\langle q3\rangle}.
\ee
The first and the last integrals can only depend on the momenta $4$ and $1$ respectively, and then vanish as causing a $\langle qq\rangle=0$ contraction. 

The second integral vanishes after a shift. First, using $l+1=l-4-2-3$ we rewrite it as
\be
\int \frac{d^4l}{(2\pi)^4}\frac{\langle q|l+1|2]\langle q| (l-4) \circ (2+3) |q\rangle}{(l+1)^2(l-4)^2\langle 41\rangle\langle 43\rangle\langle q1\rangle\langle q2\rangle\langle q3\rangle},
\ee
to make it clear that it is at most quadratically divergent. A good shift is then $l\to l-1$, which renders the integral
\be
\int \frac{d^4l}{(2\pi)^4}\frac{\langle q|l|2]\langle q| l \circ (2+3) |q\rangle}{l^2(l+2+3)^2\langle 41\rangle\langle 43\rangle\langle q1\rangle\langle q2\rangle\langle q3\rangle}.
\ee
This vanishes by the already familiar argument.

We thus need to compute the shift. The relevant integrand is
$$ \frac{\langle q|l+1|2]\langle q| (l-4) \circ (2+3) |q\rangle}{(l+1)^2(l-4)^2}.$$
The linear part of the shift is
\be
-\im \lim_{l\to \infty} \int \frac{d\Omega}{(2\pi)^4} 1_\mu l^\mu \frac{\langle q|l+1|2]\langle q| (l-4) \circ (2+3) |q\rangle}{l^2}\left( 1+2 \frac{ (l\cdot 4)-(l\cdot 1)}{l^2}\right).
\ee
The quadratic part in $l$ gives the following two terms
$$ - \frac{\im}{32\pi^2} (-\langle q|1|2]\langle q| 4 \circ (2+3) |q\rangle + \langle q|1|2]\langle q| 1 \circ (2+3) |q\rangle)= - \frac{\im}{32\pi^2} \langle q|1|2]\langle q| (1-4) \circ (2+3) |q\rangle .$$
Using $2+3=-(1+4)$ this can be rewritten as 
$$  \frac{2\im}{32\pi^2} \langle q|1|2]\langle q| 1\circ 4 |q\rangle .$$
The quartic part in $l$ gives, after simplification
$$ - \frac{\im}{3\cdot 32\pi^2} ( 3  \langle q| 1|2] -   \langle q| 4|2])\langle q| 1\circ 4|q\rangle.$$
Thus, overall, the linear part of the shift is
\be
\frac{\im}{ 32\pi^2}   \langle q| 1|2]\langle q| 1\circ 4|q\rangle + \frac{\im}{3\cdot 32\pi^2}   \langle q| 4|2]\langle q| 1\circ 4|q\rangle.
\ee

The quadratic part of the shift is
\be
\frac{\im}{2} \lim_{l\to\infty}  \int \frac{d\Omega}{(2\pi)^4} 1_{\mu}1_{\nu}l^{\mu}l^2\frac{\partial}{\partial{l_{\nu}}}\frac{\langle q|l+1|2]\langle q| (l-4) \circ (2+3) |q\rangle}{(l+1)^2(l-4)^2}.
\ee
When the derivative hits the denominator we get
$$ -\frac{2\im}{3\cdot 32\pi^2}  \langle q|1|2]\langle q| 1\circ (2+3) |q\rangle.$$
When the derivative hits the numerator we get two equal terms with the result
$$ \frac{\im}{32\pi^2} \langle q|1|2]\langle q| 1 \circ (2+3) |q\rangle $$
Overall, this gives for the quadratic part of the shift
\be
\frac{\im}{3\cdot 32\pi^2} \langle q|1|2]\langle q| 1 \circ (2+3) |q\rangle=-
\frac{\im}{3\cdot 32\pi^2}  \langle q|1|2]\langle q| 1\circ 4 |q\rangle.
\ee

Combining the linear and quadratic parts of the shift we get the full shift
\be
\frac{\im}{3\cdot 32\pi^2} (2 \langle q|1|2] + \langle q| 4|2]) \langle q| 1\circ 4|q\rangle = \frac{\im}{3\cdot 32\pi^2}  \langle q| 1-3|2] \langle q| 1\circ 4|q\rangle,
\ee
where we wrote $1+4=-2-3$. 
The integral is minus the shift. This gives the following result for the $\mathcal{A}_2$
\be
\mathcal{A}_2 = \frac{1}{12\cdot 32\pi^2}  \frac{\langle q| (3-1)\circ 2 |q\rangle  \langle q| 1\circ 4|q\rangle}{ \langle 14\rangle\langle 43\rangle\langle q1\rangle\langle q2\rangle^2\langle q3\rangle},
\ee
where we wrote the result in form of (\ref{res-m1}) by multiplying the numerator and denominator by $\langle 2q\rangle$.

\subsection{Collecting the results}

Collecting the results for $\mathcal{A}_{1,2,3}$ we get for a single box diagram
\be
12 \cdot 32\pi^2 \mathcal{A}= -\frac{[12][23]}{\langle 23\rangle \langle 43\rangle} \frac{\langle q2 \rangle \langle q3 \rangle}{\langle q1 \rangle\langle q4 \rangle} + \frac{[14][23]}{\langle 23\rangle \langle 43\rangle} \frac{ \langle q3 \rangle}{\langle q1 \rangle} \\ \nonumber
-\frac{[12][14]}{\langle 14\rangle \langle 43\rangle} \frac{\langle q1 \rangle \langle q4 \rangle}{\langle q2 \rangle\langle q3 \rangle} - \frac{[14][23]}{\langle 14\rangle \langle 43\rangle} \frac{ \langle q4 \rangle}{\langle q2 \rangle} + \frac{[12]^2}{\langle 12\rangle \langle 43\rangle} \frac{\langle q1 \rangle \langle q2 \rangle}{\langle q3 \rangle\langle q4 \rangle} .
\ee
As a check, we note that the replacement $1\leftrightarrow 2$ gives the same result as $3\leftrightarrow 4$. This is because the amplitude is symmetric under the simultaneous transformation $1\leftrightarrow 2$ and $3\leftrightarrow 4$. 

The above amplitude is not manifestly invariant under the simultaneous $2\leftrightarrow 4$ and $3\leftrightarrow 1$ exchange. To rectify this, we will rewrite the two terms that have single powers of $q$ in the numerator and denominator. We have
\be
\frac{[14][23]}{\langle 23\rangle \langle 43\rangle} \frac{ \langle q3 \rangle}{\langle q1 \rangle} - \frac{[14][23]}{\langle 14\rangle \langle 43\rangle} \frac{ \langle q4 \rangle}{\langle q2 \rangle} =\frac{[14][23]}{\langle 14\rangle \langle 23\rangle \langle 43\rangle \langle q1 \rangle \langle q2 \rangle} \left( \langle 14\rangle \langle q2 \rangle \langle q3 \rangle-\langle 23\rangle \langle q1 \rangle \langle q4 \rangle\right).
\ee
We now use the following Schouten identities for the two terms in the brackets
\be
\langle q4 \rangle \langle 23 \rangle = \langle q2 \rangle \langle 43 \rangle -\langle q3 \rangle \langle 42 \rangle , \qquad
\langle q2 \rangle \langle 14 \rangle = \langle q1 \rangle \langle 24 \rangle -\langle q4 \rangle \langle 21 \rangle ,
\ee
which gives
\be
\langle 14\rangle \langle q2 \rangle \langle q3 \rangle-\langle 23\rangle \langle q1 \rangle \langle q4 \rangle  = \langle q1 \rangle \langle q2  \rangle \langle 34\rangle + \langle q3 \rangle \langle q4  \rangle \langle 12\rangle.
\ee
This means that the amplitude can be written as
\be
12 \cdot 32\pi^2 \mathcal{A}= \frac{[12][23]}{\langle 23\rangle \langle 34\rangle} \frac{\langle q2 \rangle \langle q3 \rangle}{\langle q1 \rangle\langle q4 \rangle} +\frac{[12][14]}{\langle 14\rangle \langle 34\rangle} \frac{\langle q1 \rangle \langle q4 \rangle}{\langle q2 \rangle\langle q3 \rangle}  \\ \nonumber - \frac{[12]^2}{\langle 12\rangle \langle 34\rangle} \frac{\langle q1 \rangle \langle q2 \rangle}{\langle q3 \rangle\langle q4 \rangle}
- \frac{[34]^2}{ \langle 12\rangle \langle 34\rangle} \frac{\langle q3 \rangle\langle q4 \rangle}{\langle q1 \rangle \langle q2 \rangle}- \frac{[14][23]}{\langle 14\rangle \langle 23\rangle},
\ee
where we also used $[23]=[34]\langle 14\rangle/\langle 12\rangle$ and $[14]=[34]\langle 23\rangle/\langle 12\rangle$ in the second term on the second line. Note that the last term is $q$-independent and thus gauge invariant. 

The invariances are now manifest. The gauge-invariant term is manifestly invariant under $1\leftrightarrow 2$ and $3\leftrightarrow 4$ as well as $2\leftrightarrow 4$ and $3\leftrightarrow 1$ symmetry. The terms in the first line go one into another under both of these symmetries, and hence also invariant. The first two terms in the second line are separately invariant under $1\leftrightarrow 2$ and $3\leftrightarrow 4$, while under $2\leftrightarrow 4$ and $3\leftrightarrow 1$ they go into each other. 

\subsection{Full amplitude}

We can now obtain the full amplitude by taking the above $\mathcal{A}$ and adding to it its $1\leftrightarrow 2$ as well as $2\leftrightarrow 3$ permutations. It is then easy to check that all $q$ dependent terms cancel pairwise as the result of momentum conservation identities. What remains is only the sum of the corresponding $q$-independent terms
\be
-12 \cdot 32\pi^2 \mathcal{A}_{one-loop}(1^+,2^+,3^+,4^+) =  \frac{[14][23]}{\langle 14\rangle \langle 23\rangle} + \frac{[24][13]}{\langle 24\rangle \langle 13\rangle}+ \frac{[14][32]}{\langle 14\rangle \langle 32\rangle}.
\ee
The second term here is the same as the other two by momentum conservation and we finally get
\be
\mathcal{A}_{one-loop}(1^+,2^+,3^+,4^+) =-\frac{1}{128\pi^2}  \frac{[12][34]}{\langle 12\rangle \langle 34\rangle},
\ee
where we rewrote the answer in the more familiar form using momentum conservation identities. This is the expected answer (\ref{amplitude-intr}).

\section*{Acknowledgement} KK thanks Zvi Bern for a discussion on the subject of this paper. 

\section*{Appendix}

In this Appendix we spell out the 2-component spinor version of the Feynman rules used in this paper. Our 2-component spinor conventions are explained e.g. in the Appendix of \cite{Krasnov:2016emc}. All computations in the main text are done for massless QED. As we explain below, they extend with minimal modifications to the case of self-dual YM theory. The last subsection spells out the rules used for extracting the shifts.

\subsection*{Massless QED Feynman rules}

The standard reference for the 2-component spinor formalism is \cite{Dreiner:2008tw}. Our conventions are analogous, the main change being that we use the relativists capital Latin letters (primed and unprimed) for the spinor indices rather than the particle physics dotted and undotted letters. In 2-component spinor notations, the fermionic part of the Lagrangian of the massless QED is
\be
{\cal L} = \im (\psi^\dagger)^{M'} (\partial_{M'}{}^M + \im A_{M'}{}^{M}) \psi_M.
\ee
Here $\psi_M$ is a 2-component (unprimed) spinor (i.e. a Weyl spinor), and $(\psi^\dagger)_{M'}$ is its Hermitian conjugate (primed) 2-component spinor. The object $A_{M'}{}^M$ is the spinor version of the electromagnetic potential $A_\mu$. This Lagrangian is real modulo integration by parts and a surface term. 

The resulting spinor propagator is
\be\label{prop-qed}
\langle \psi_{M}(-k) (\psi^\dagger)^{M'}(k) \rangle = \frac{2}{\im k^2} k_M{}^{M'},
\ee
where the factor of 2 originates from this factor in (\ref{l2-ident}). This is depicted by an ordered line from the primed spinor to the unprimed. 

The vertex factor is 
\be\label{vertex-qed}
\langle \psi^\dagger_{M'} \psi^M A_{N'}{}^N \rangle = \im \epsilon_{M'N'}\epsilon^{MN},
\ee
times the usual $\delta$-function for the momentum conservation. In writing the expressions for the amplitudes such as (\ref{tri}) or (\ref{boxamp}) we omitted the inessential numerical factors of $2$ for each propagator, as well as the factors of the imaginary unit for each vertex and each propagator. So, the final expressions for all the amplitudes must be corrected by these numerical and phase factors. 

\subsection*{Self-dual YM Feynman rules}

We use the description of self-dual YM and its gauge-fixing as explained in \cite{Krasnov:2016emc}, see in particular formulas (13), (14) of this references for the 2-component spinor version of the gauge-fixed Lagrangian. The kinetic term is given by
\be
{\cal L}^{(2)} = b^{a \, NM} \partial_M{}^{M'} a^a_{NM'}.
\ee
Here $a$ is the Lie algebra index, and $b^{a\, NM}$ and $a^a_{MM'}$ are the perturbations of the self-dual auxiliary field and the connection respectively. As is explained in \cite{Krasnov:2016emc}, after gauge-fixing the auxiliary field $b^{a\, NM}$ acquires an additional component and is no longer symmetric in its two spinor indices. Thus, the propagator of this theory is
\be
\langle b_A^{a\, M}(-k) a^b_{BM'}(k)\rangle = \frac{2}{k^2} \delta^{ab} \epsilon_{AB} k_{M'}{}^{M}.
\ee
The factor of 2 is again from (\ref{l2-ident}). There is no factor of the imaginary unit in this propagator because this factor is not present in the Lagrangian. Thus, apart from the different phase factor, the only difference in the propagator of the self-dual YM from (\ref{prop-qed}) is that there is an extra Kronecker delta for the colour, as well an extra factor of the spinor metric for the additional unprimed index present in the SDYM as compared to the case of QED. We are not interested in the phase factors in this paper, and so we did not attempt to get the sign of the propagator right. 

The interaction term in the SDYM Lagrangian is
\be
2 f^{abc} b^{a\, MN} a^b_M{}^{M'} a^c_{NM'},
\ee
where $f^{abc}$ is the Lie algebra structure constant. This results in the following vertex factor
\be
\langle b^{a\, MN} a^b_{AM'} a^c_{BN'}\rangle = 2\im f^{abc} \epsilon_{M'N'} \epsilon_A{}^M \epsilon_B{}^N.
\ee
This should be compared with the QED vertex (\ref{vertex-qed}). We see that the vertex factor of SDYM has an additional colour factor, as well as an additional factor of the spinor metric for the unprimed index. Thus, one can see that the only change one has to do in the computation of the SDYM box diagram (apart from introducing the colour factors) is to introduce an additional set of spinor metrics for the unprimed spinor index. This shows, directly from the Feynman rules, that the SDYM result is a multiple of the massless QED result. 

\subsection*{Extracting the shift}

Let us consider the difference of two one dimensional quadratically divergent integrals of the form 
\be 
J(b)-J(0)\equiv\int_{-\infty}^{+\infty}dy [f(y+b)-f(y)],
\ee 
where $f(\pm\infty)=d_{\pm}$ are finite constants. If the integral were convergent, it is easy to see that $J(b)-J(0)$ would vanish by a quick change of integration variables $y+b\rightarrow y'$. In the present case, however, let us Taylor expand $f(y+b)$ in powers of $b$. Note that second and higher derivatives of $f(y)$ vanishes at $y=\pm\infty$.
\be 
J(b)=\int_{-\infty}^{+\infty}dy[f(y)+bf'(y)+\frac{1}{2}b^2f''(y)+...]\nonumber
\ee 
\be 
=J(0)+b(f(+\infty)-f(-\infty))+\frac{1}{2}b^2(f'(+\infty)-f'(-\infty))
\ee 
We see that the difference gives a finite value in this case.
\be 
J(b)-J(0)=b(f(+\infty)-f(-\infty))+\frac{1}{2}b^2(f'(+\infty)-f'(-\infty))
\ee 
Now replacing this with an integral over a 4-dimensional momentum, we find analogously
\be 
\label{limit}
\int \frac{d^4l}{(2\pi)^4}[f(l+b)-f(l)]=\im \lim_{l\to\infty} \int \frac{d\Omega}{(2\pi)^4} b_\mu l^\mu l^2 f(l)+\frac{\im}{2} \lim_{l\to\infty} \int \frac{d\Omega}{(2\pi)^4} b_\mu b_\nu l^\mu l^2 \frac{\partial}{\partial l_\nu} f(l).
\ee
Here $\im$ is the imaginary unit that comes from the analytic continuation to the Euclidean signature, and $d\Omega$ is the area element on the unit sphere of the directions of vector $l^\mu$.

\end{fmffile}

\end{document}